\documentclass[%
 reprint,
 amsmath,amssymb,
 aps,
]{revtex4-2}

\usepackage{graphicx}
\usepackage{dcolumn}
\usepackage{bm}

\usepackage[acronyms]{glossaries}
\newacronym{aligo}{aLIGO}{Advanced LIGO}
\newacronym{gw}{GW}{gravitational wave}
\newacronym{cbc}{CBC}{compact binary coalescence}
\newacronym{cw}{CW}{continuous wave}
\newacronym{sgwb}{SGWB}{stochastic gravitational-wave background}
\newacronym{fea}{FEA}{finite-element analysis}
\newacronym{gnd}{GND}{ground}
\newacronym{top}{TOP}{top}
\newacronym{uim}{UIM}{upper intermediate mass}
\newacronym{pum}{PUM}{penultimate mass}
\newacronym{tst}{TM}{test mass}
\newacronym{ssm}{SSM}{state-space model}
\newacronym[plural=DoF, firstplural=Degrees of Freedom (DoF)]{dof}{DoF}{Degree of Freedom}
\newacronym{swg}{SWG}{Seismic \& Suspensions Working Group of the LIGO Scientific Collaboration}
\newacronym{lho}{LHO}{LIGO Hanford Observatory}
\newacronym{pem}{PEM}{physical environment monitor}
\newacronym{msup}{MSUP}{Mode Superposition}

\begin{document}

\preprint{APS/123-QED}

\title{Identification of Resonant Frequencies in LIGO-like Suspension with Finite-Element Modeling}

\author{Orion Sauter}
\email{orionsauter@ufl.edu}
\author{Ninad Bhagwat}
\email{ninadgbhagwat@gmail.com}
\author{John Conklin}
\email{jwconklin@ufl.edu}
\author{D.B. Tanner}
\email{tanner@phys.ufl.edu}

\affiliation{University of Florida}

\date{\today}

\begin{abstract}
Following the upgrades to \gls{aligo}, measurements were made of the detector suspensions' frequency response characteristics. While most resonant frequencies could be identified with simple mechanical models, such as the fiber vibration modes, some were unexplained. Using a finite element model of the quadruple pendulum suspension, we search for and identify lines from unknown sources. The present work focuses on two resonant lines observed in the Upper Intermediate Mass as examples of this technique. Our simulations suggest a common source for these lines, which could be accounted for in a redesign. By modeling these response frequencies, we can examine the motion of individual components, and suggest methods to reduce their amplitude, alter their frequency, or eliminate them in future gravitational wave detector designs.
\end{abstract}

\maketitle


\section{Introduction}
The \gls{aligo} detectors were designed to detect \glspl{gw} with frequencies in the range 10 Hz to 7000 Hz \cite{aLIGO}. The detectors use a Fabry-Perot interferometer to measure the change in differential arm length created by passing gravitational waves. However, the length changes induced by these waves are $O(10^{-21})$ in strain, and can easily be covered by noise. To achieve the necessary stability, the detectors' suspensions use quadruple pendulums to suspend the test masses, which serve as the end mirrors for the interferometer \cite{quad, quad_2012}. Some noise sources are transient, and the affected time spans can be vetoed without much reduction in SNR \cite{vetos}. However, more persistent noise sources affecting a small range of frequencies, such as mechanical resonances and electrical coupling, can degrade the performance of both \gls{sgwb} and \gls{cw} searches when the signal frequency passes through the frequency of the disturbed band. When the noise spectrum is plotted on a semilog scale over many decades, these noise bands appear as single vertical lines; hence, we call them ``lines.'' By identifying and resolving these noise sources, we can improve our detection capabilities. Due to the complexity of the detectors, it is not always clear what causes a given spectral line in the strain sensitivity. If we can find a similar line in a simplified surrogate model, we may be able to identify the source.

\Gls{fea} offers a high-fidelity simulation which tracks all components' Degrees of Freedom (DoF), including internal models of mechanical components. As an initial target, we choose the \gls{aligo} quadruple pendulum suspensions. These consist of four stages, which descend from the top-level supports: maraging steel blade springs on the top supports connect with two 45 cm long steel wires to the \gls{top} mass weighing 20 kg. Blade springs on the \gls{top} connect with four 31 cm long steel wires to the \gls{uim} weighing 20 kg. Blade springs on the \gls{uim} connect with four 34 cm long steel wires to the \gls{pum} weighing 40 kg. Fused silica fibers connect the \gls{pum} directly to the \gls{tst} weighing 40 kg \cite{quad_2012}. Resonances from the upper masses can couple into the \gls{tst} displacement, which is measured by the interferometer.

In Section \ref{sec:spec_lines} we outline the problem of spectral lines. Section \ref{sec:fea} describes the model we used to simulate the detector, and Section \ref{sec:dof} outlines the conversion from the \gls{fea} outputs to a six DoF system used to describe the suspensions. Sections \ref{sec:tf}, \ref{sec:line}, and \ref{sec:line2} compare that output to \gls{lho} measurements via a transfer function.

\section{Unknown Spectral Lines}
\label{sec:spec_lines}
The \gls{aligo} Detector Characterization group maintains \glspl{ssm} of the \gls{aligo} pendulums for simulation and testing. These models each comprise four matrices which define the time evolution from inputs -- applied forces and torques -- to outputs -- displacements and rotations:
\begin{eqnarray}
    \dot{\mathbf{x}} =& A\mathbf{x} + B\mathbf{u}\\
    \mathbf{y} =& C\mathbf{x} + D\mathbf{u}
\end{eqnarray}
where $\mathbf{u}$ are inputs, $\mathbf{y}$ outputs, and $\mathbf{x}$ states of the system. For second-order systems, the states can be split into a set of first-order states, followed by their derivatives, i.e.
\begin{equation}
    \mathbf{X} = \begin{bmatrix}
        \mathbf{x}\\ \mathbf{\dot{x}}
    \end{bmatrix}.
\end{equation}
This results in the matrices having a block form, as in the initial version of the \gls{ssm}, based on a theoretical model of the pendulum, which used rigid body masses, and linearized springs \cite{ssm}. In this case, the state variables corresponded directly to the six DoF of each of the stages. However, this model was found to be missing higher-frequency features present in the detector's true response behavior. In light of this, a more detailed \gls{ssm} was developed, introducing numerous filters and measurements from the detectors as-built \cite{ssm2020}. Unlike the previous model in which states correspond to the individual stages' DoF, the 1483 states of the 2020 \gls{ssm} are the result of a chain of calculations, making it difficult to determine the source of any given output. 
After the second \gls{aligo} observing run (O2) measurements were made of narrow spectral lines in the detectors \cite{o2_spec}. By using \glspl{pem}, we can identify lines due to local disturbances \cite{pem}. Other lines can be attributed to known mechanical features, such as the vibrational modes of the suspensions \cite{o2_sens}. However, some lines could not be identified. Without knowing the sources of these lines, we have no way to suppress or alter them.

\section{Finite-Element Model}
\label{sec:fea}
\begin{figure}[htb]
    \centering
    \includegraphics[width=\columnwidth]{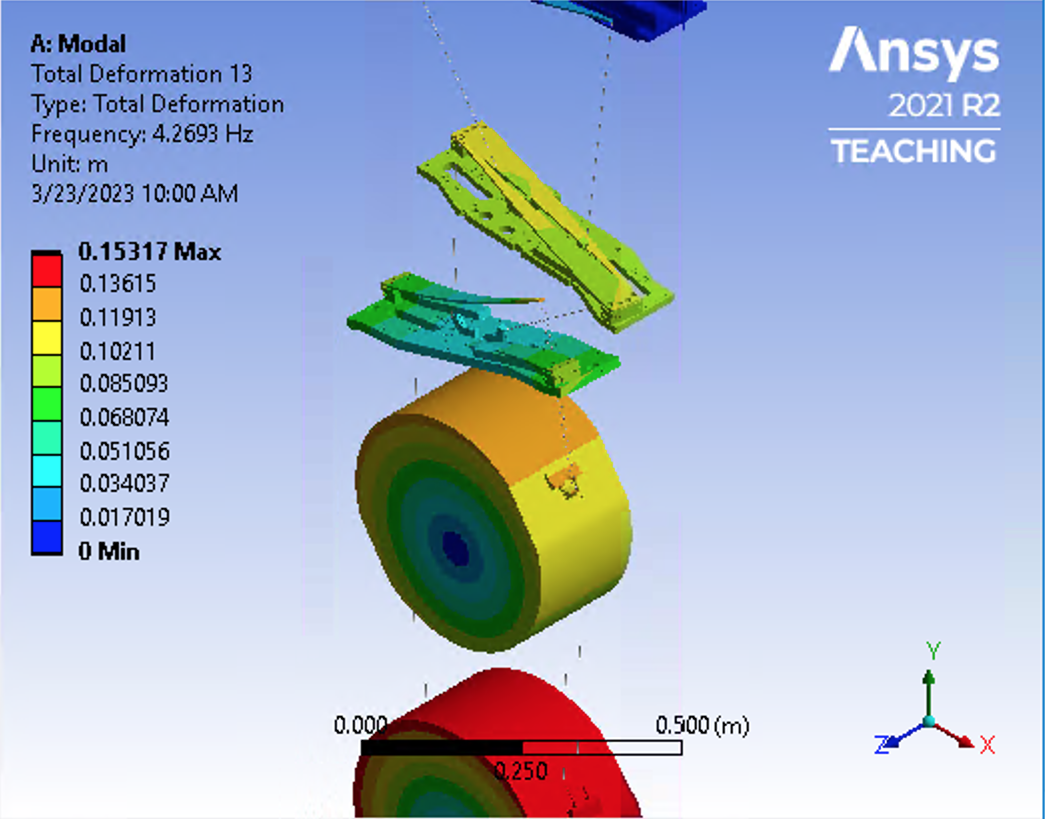}
    \caption{Mode including bending of blade springs.}
    \label{fig:blade_mode}
\end{figure}

\Gls{fea} is a common technique in engineering and design to analyze the stresses within a large structure under applied forces. The University of Glasgow has developed a finite-element model of the \gls{aligo} quadruple pendulums using the ANSYS software package \cite{glas_quad}. The model simplifies many of the details of the physical system, such as replacing screw connections with bonded contacts. Glasgow used this model to analyze the modes involving the blade springs that support the weight of the test masses (Figure \ref{fig:blade_mode}).

Although the motion of the stages is driven by eigenmodes calculated from modal analysis, motion response is better calculated with transient time series analysis. In this method a non-periodic, non-harmonic time dependent external load is applied to the structure and the resultant displacement, strain, stress, and reaction forces are computed using the following governing equation:
\begin{equation}
M \ddot{\mathbf{u}}(t) + Z \dot{\mathbf{u}}(t) + K \mathbf{u}(t) = \mathbf{F}(t),
\end{equation}
where $M$ is the mass matrix, $Z$ is the damping matrix, $K$ is the stiffness matrix, $F(t)$ is the applied load, and $\mathbf{u}(t)$ is the nodal displacement.
 
 There are various finite element methods to solve the above governing equation. Among these are the full method, \gls{msup} method etc. The \gls{msup} method first computes the mode shapes using modal analysis and then combines them to obtain the higher frequency modes. The \gls{msup} method is faster and more efficient but only supports linearity. The full method uses the whole matrices to calculate the dynamic behavior of the system. Unlike the \gls{msup} method, the full method supports non-linearity and produces accurate results for large displacements. Since our model involves high stresses due to the interaction between the blade springs and large stage masses, we use the full method to ensure accurate simulation.
 
 Unfortunately, this introduces further complications, as we must apply gravity to have a restoring force, and the flat springs used in the Glasgow model are not able to support the 40 kg masses that make up the lowest two stages. In the physical system, the blades are made with an upward curvature. When the blades are loaded with the test masses, the upward curvature is countered by the weight acting downward due to gravity. At equilibrium, weight-induced stress causes the blades to become flat. By simulating these blade springs in isolation, we can generate the induced stresses from the bending force, then import them onto the flat springs in the full model.

\begin{figure}[htb]
    \centering
    \includegraphics[width=\columnwidth]{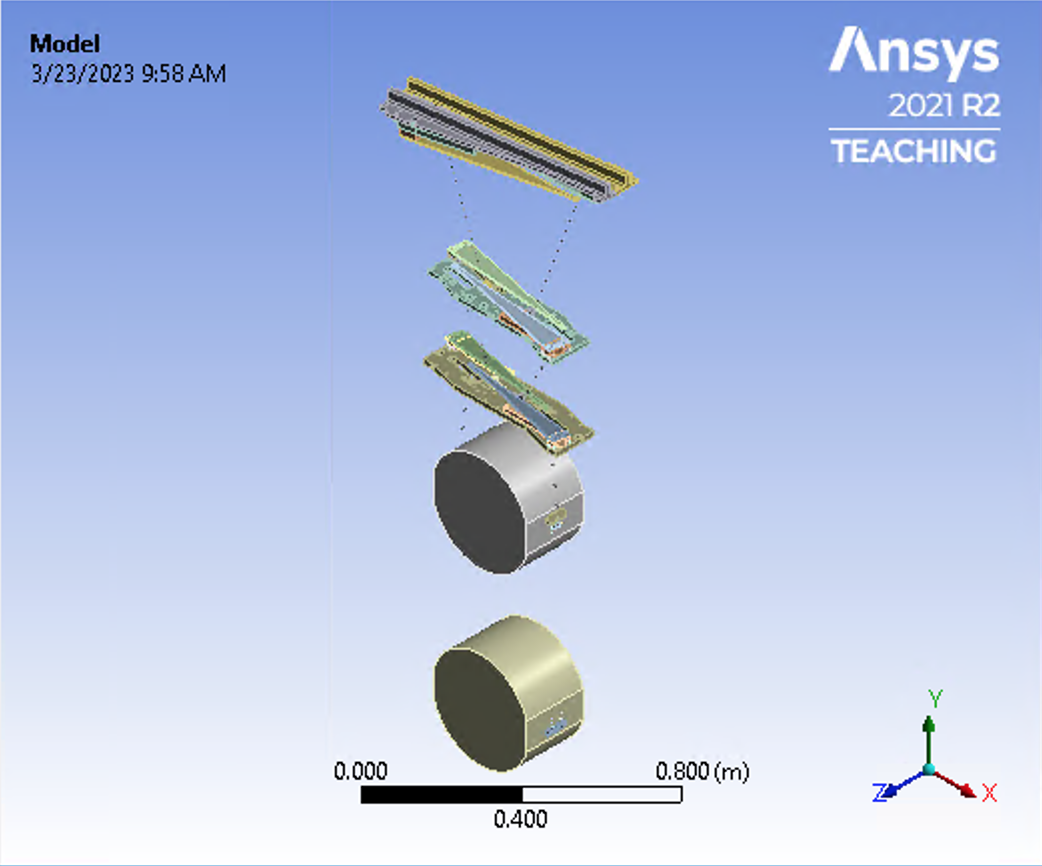}
    \caption{University of Glasgow finite-element model of the \gls{aligo} quadruple pendulum.}
    \label{fig:glas_quad}
\end{figure}

\section{Six Degree of Freedom Aggregation}
\label{sec:dof}
For simplicity, the position of each stage is typically expressed in terms of longitudinal, vertical, and transverse displacements, and roll, yaw, and pitch angles. In Figure \ref{fig:glas_quad}, these correspond to displacements and rotations in z, y, and x, respectively. However, the ANSYS results contain information for the individual nodes that make up the model. To convert these, for each component we find the centroid $\mathbf{c}$, and then for each node the relative position $\mathbf{r} = \mathbf{p}-\mathbf{c}$. Each node has an initial position expressed as a vector from the origin, $\mathbf{p}^{\,\mathrm{loc}}$, and a displacement expressed as a vector from the initial position, $\mathbf{p}^{\,\mathrm{disp}}$. Using these properties, we can define the six DoF, which we average over the component's nodes:
\begin{eqnarray}
L =& p^{\,\mathrm{disp}}_z\\
V =& p^{\,\mathrm{disp}}_y\\
T =& p^{\,\mathrm{disp}}_x\\
R =& \frac{1}{2} \left(\arctan{\left(p^{\,\mathrm{disp}}_y \big / r^{\,\mathrm{loc}}_z\right)} + \arctan{\left(p^{\,\mathrm{disp}}_z \big /r^{\,\mathrm{loc}}_y\right)}\right)\\
Y =& \frac{1}{2} \left(\arctan{\left(p^{\,\mathrm{disp}}_x \big / r^{\,\mathrm{loc}}_z\right)} + \arctan{\left(p^{\,\mathrm{disp}}_z \big /r^{\,\mathrm{loc}}_x\right)}\right)\\
P =& \frac{1}{2} \left(\arctan{\left(p^{\,\mathrm{disp}}_y \big / r^{\,\mathrm{loc}}_x\right)} + \arctan{\left(p^{\,\mathrm{disp}}_x \big /r^{\,\mathrm{loc}}_y\right)}\right)
\end{eqnarray}

\section{Transfer Function Estimation}
\label{sec:tf}
The measurements made by the \gls{aligo} \gls{swg} are in the form of transfer functions relating applied forces and torques to displacements and rotations for the DoF of the stages of the pendulum. For comparison, we need to generate similar transfer functions for our model.

For a given input stage and DoF, we apply Gaussian noise as a force along the DoF on all nodes in the stage, with a 1 kHz sampling rate in a transient analysis. We then extract the six DoF displacements of the stages in time, and take the desired output DoF. The estimated transfer function is then the Fourier transform of the output divided by the Fourier transform of the input.

\section{Case Study: 166 Hz Line}
\label{sec:line}
\begin{figure}[htb]
    \centering
    \includegraphics[width=\columnwidth]{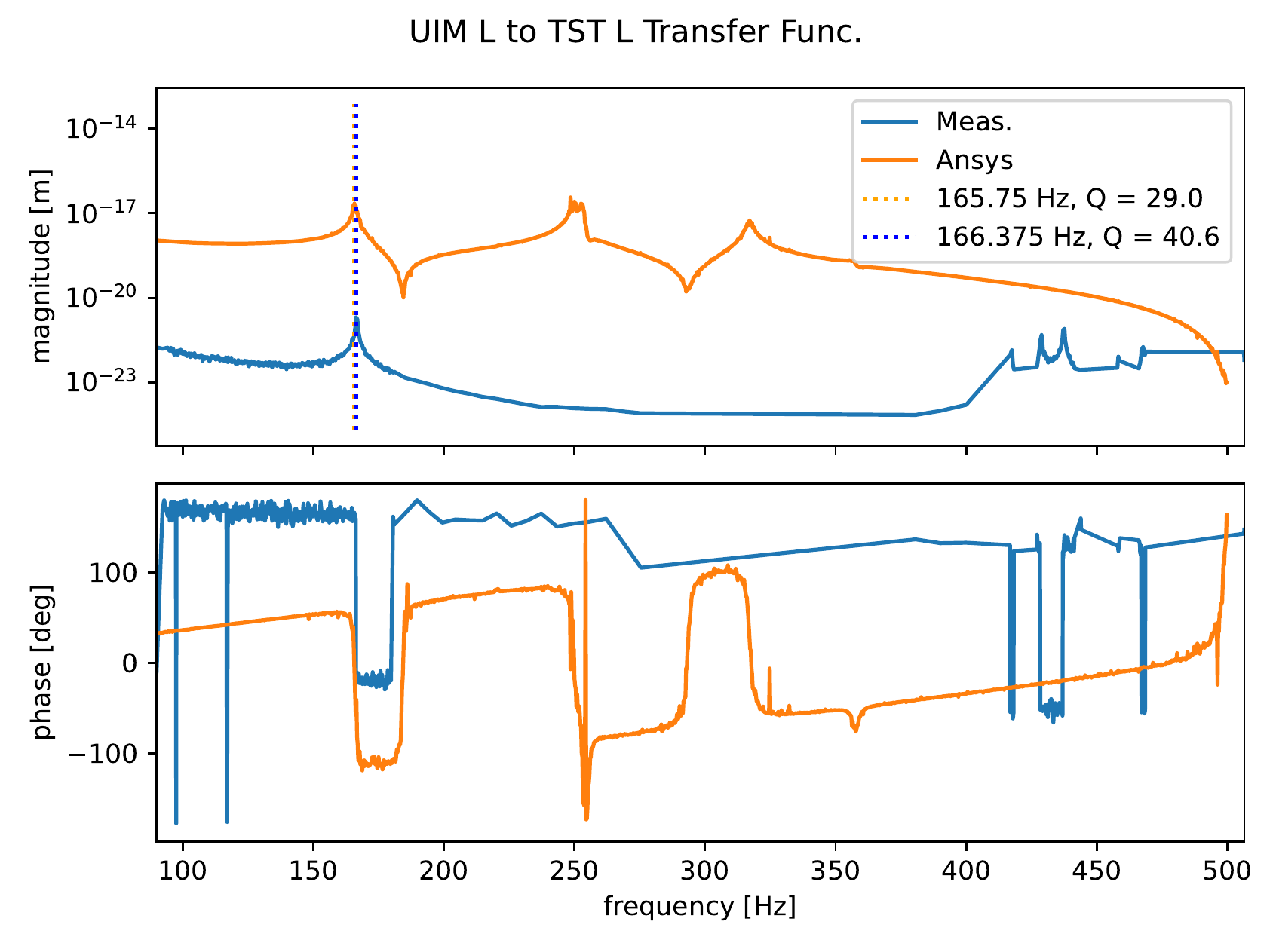}
    \caption{Measured and ANSYS-modeled transfer functions between \gls{uim} and \gls{tst} longitudinal DoF.}
    \label{fig:xfer}
\end{figure}

Among the unidentified spectral lines found following O2 was one at approximately 166 Hz, in the transfer function between the longitudinal DoF of the \gls{uim} and \gls{tst}. Despite efforts by the \gls{swg}, the source of this line could not be found. We chose this as a target study for the use of our model. Using the technique described above, we estimated the transfer function for this coupling in the ANSYS model, and show a comparison to measurements of \gls{lho} in Figure \ref{fig:xfer}. We were able to bring the resonances into better alignment by adjusting the damping coefficient used in the simulation, and reducing the Young's modulus of the steel base plate in the model by 11\%, an amount that is within 2 standard deviations of the mean for stainless steel \cite{steel}.

For each transfer function, we find the peak frequency and Q factor, calculated from estimates of the full-width at half-maximum of the curves shown in Figure \ref{fig:xfer}. The ANSYS model displays a peak at 165.75 Hz with $Q = 29.0$, while the measured peak is at 166.375 Hz with $Q = 40.6$. While bringing these peaks into agreement required modifying the model, the adjustments are small enough to hypothesize a common source for the peaks.

\begin{figure}[htb]
    \centering
    \includegraphics[width=\columnwidth]{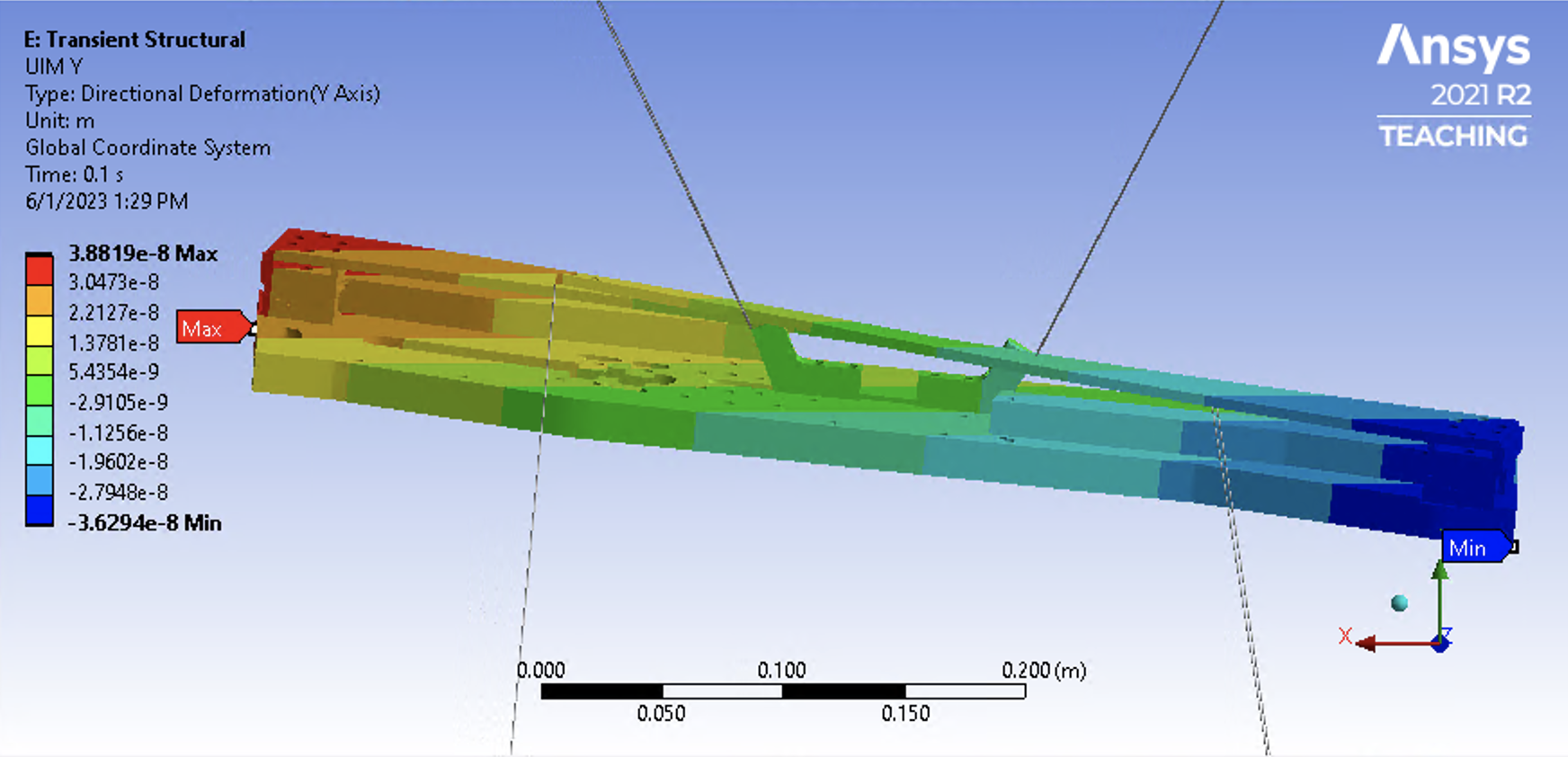}
    \caption{Vertical motion of \gls{uim} base plate under 166.375 Hz longitudinal force on the \gls{uim} nodes.}
    \label{fig:uim_tot}
\end{figure}

\begin{figure}[htb]
    \centering
    \includegraphics[width=\columnwidth]{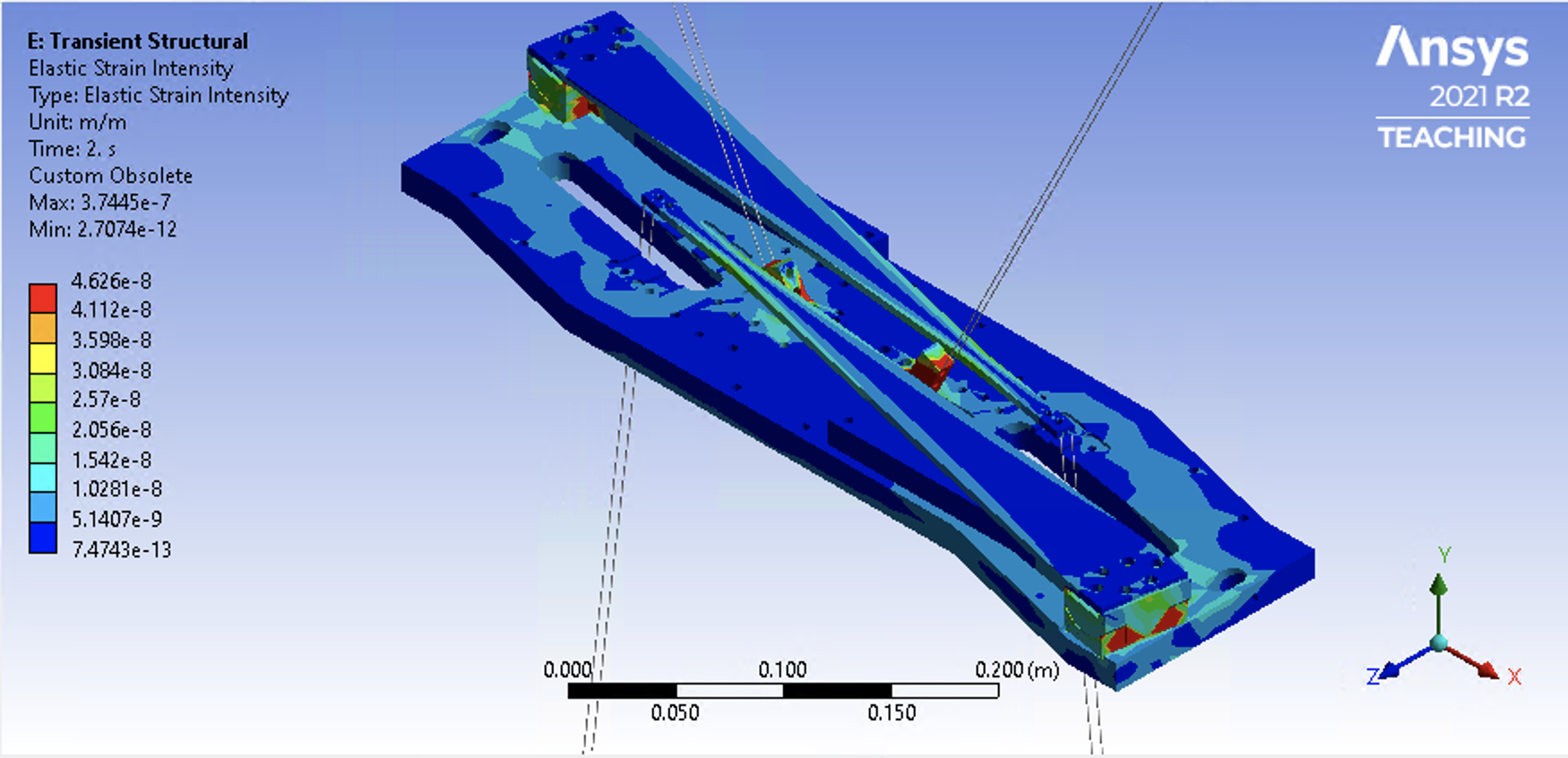}
    \caption{Elastic strain in \gls{uim} under 166.375 Hz excitation.}
    \label{fig:uim_strain}
\end{figure}

\begin{figure}[htb]
    \centering
    \includegraphics[width=\columnwidth]{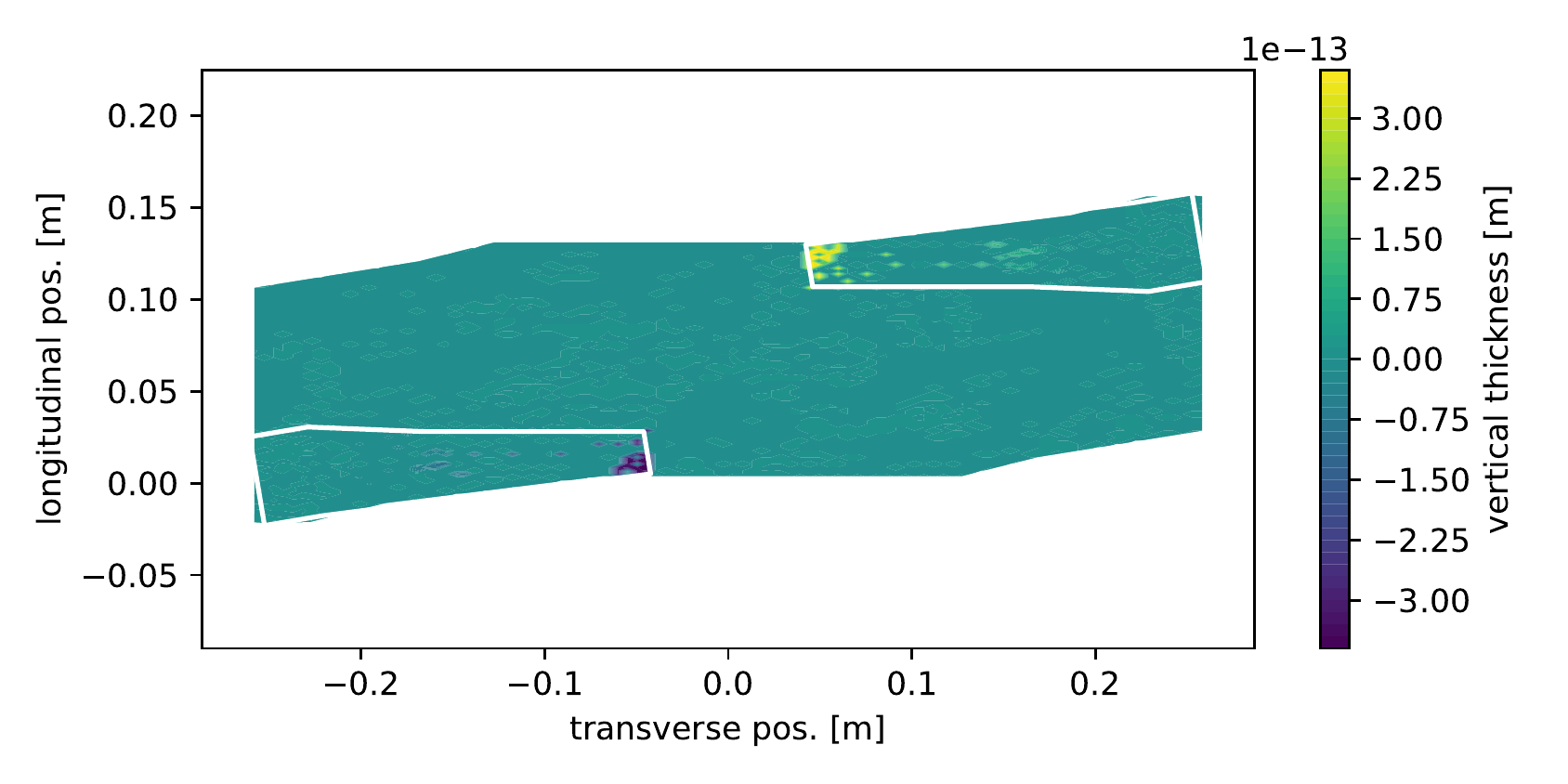}
    \caption{Change in \gls{uim} base plate thickness under force. White lines show the outlines of the spring supports. The apparent discontinuities are due to the resolution of the simulation. However, these differences are small, and it is through the stiffness of the material that they are able to have a measurable effect.}
    \label{fig:uim_def}
\end{figure}

Assuming these peaks result from similar motion of the pendulum, we can use ANSYS to determine what components are responsible for the behavior. If we drive the \gls{uim} stage's nodes longitudinally with a sinusoidal force at the resonant frequency, and determine the vertical displacement for points across the transverse axis of the \gls{uim}, we see a rocking motion shown in Figure \ref{fig:uim_tot}. However, this motion is more commonly seen at low frequency, due to the diagonal positioning of the blade springs: The tips are on the front/back center line, but the bases are fixed on opposite corners, resulting in coupling between the longitudinal and transverse DoF, as well as the vertical displacement. By looking at the strain distribution over the components of the \gls{uim} in Figure \ref{fig:uim_strain}, we see there is significant energy in the connection points between the blade springs and the base plate. We can bandpass the motion in the range 150 Hz to 200 Hz, and look at the vertical deformation over the surface of the \gls{uim} base plate (Figure \ref{fig:uim_def}).

The two regions of high-deformation in Figure \ref{fig:uim_def} correspond to the blocks connecting the blade springs to the base plate. The two connection points vibrate in opposite directions at the frequency found from the transfer functions. Reinforcing the base plate in the regions where the blade springs connect may help to reduce this motion and increase the frequency.

\section{Case Study: 431 Hz Line}
\label{sec:line2}
\begin{figure}
    \centering
    \includegraphics[width=\columnwidth]{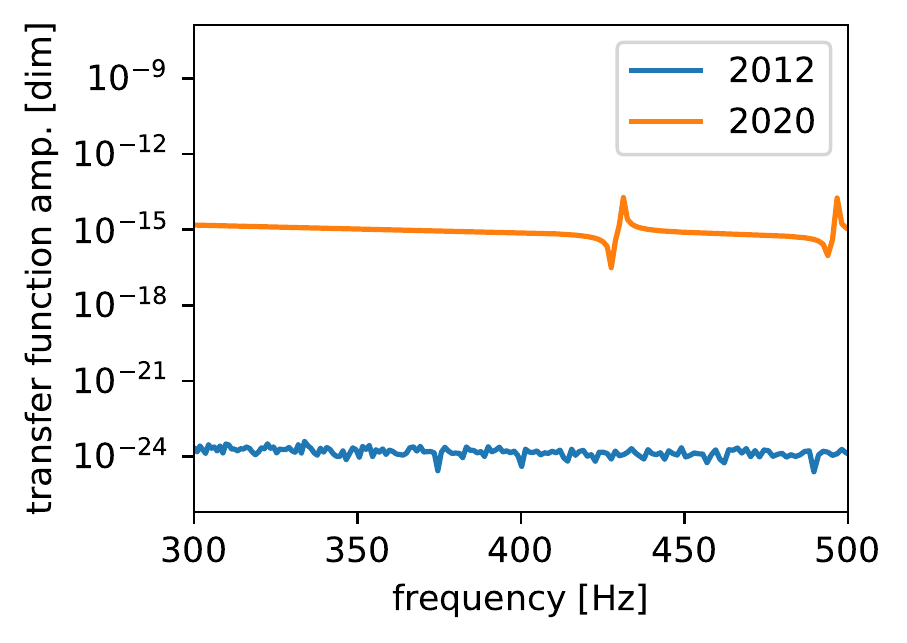}
    \caption{\gls{uim} vertical-to-\gls{tst} longitudinal transfer functions for 2012 and 2020 \glspl{ssm}. The 431 Hz line was not present in the earlier model, and the plot simply shows the noise floor for the simulation.}
    \label{fig:v2l}
\end{figure}

\begin{figure}[htb]
    \centering
    \includegraphics[width=\columnwidth]{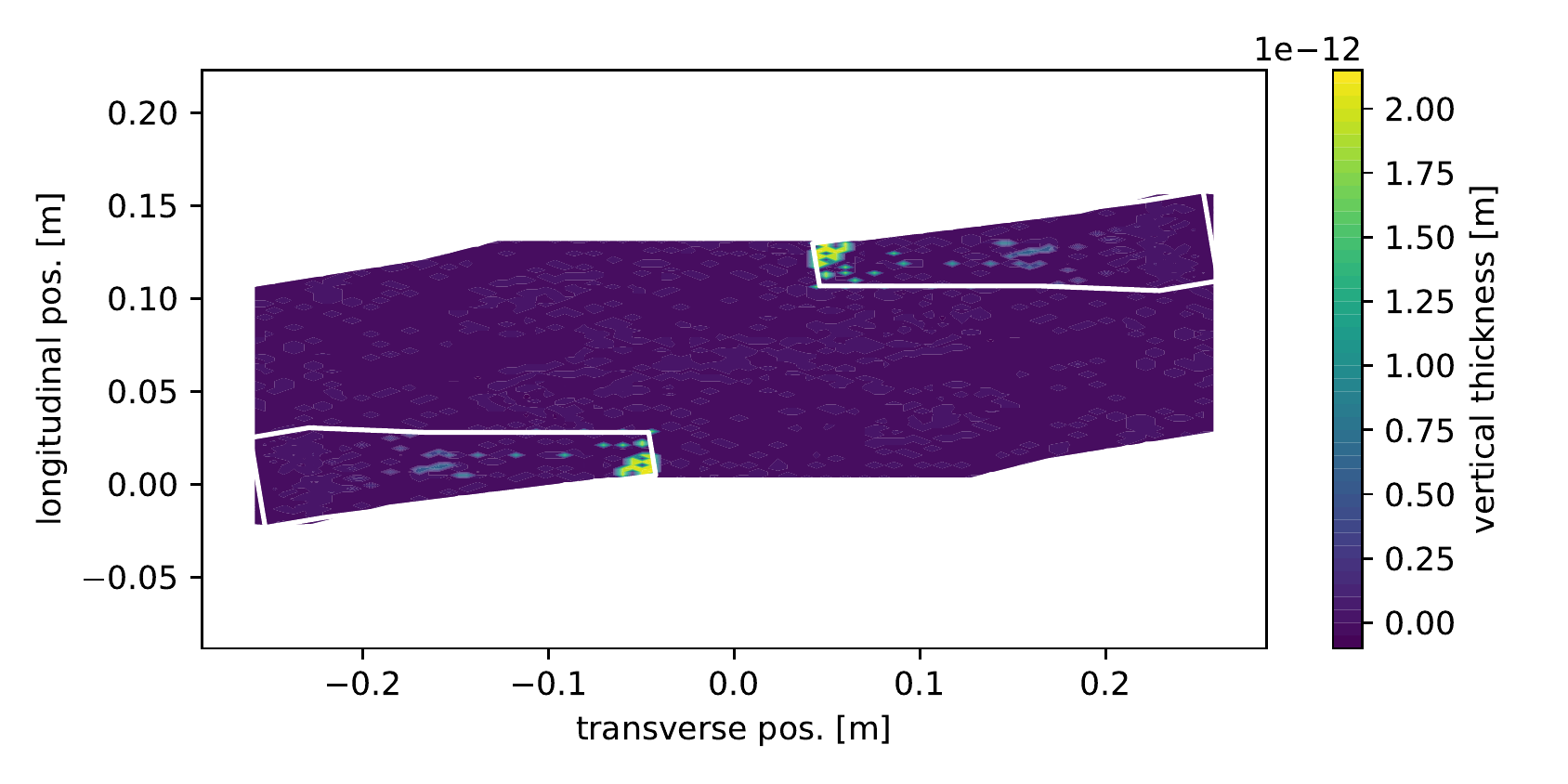}
    \caption{Change in \gls{uim} base plate thickness under force. White lines show the outlines of the spring supports.}
    \label{fig:uim_def_431}
\end{figure}

\begin{figure}
    \centering
    \includegraphics[width=\columnwidth]{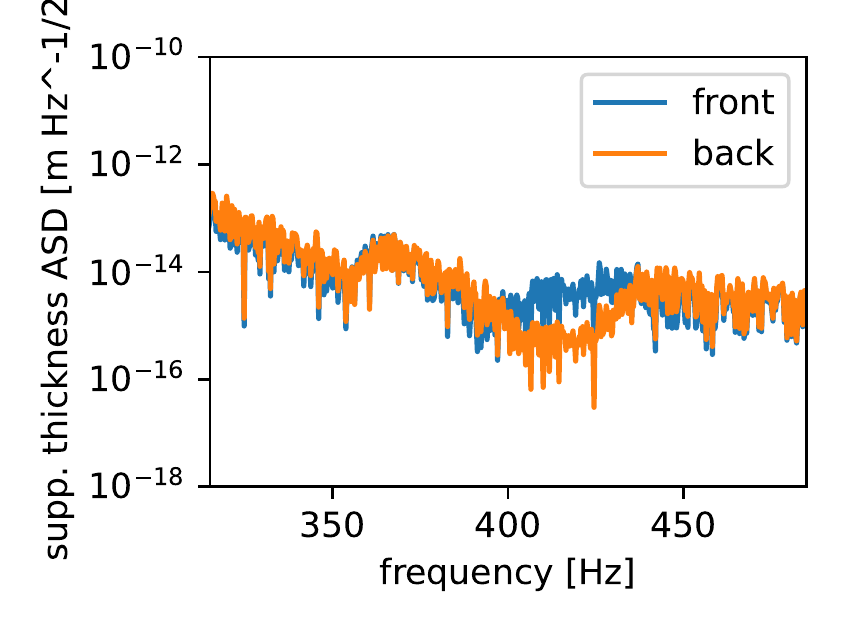}
    \caption{\gls{uim} thickness under vertical displacement.}
    \label{fig:v2lmeas}
\end{figure}

Looking at differences in response between the two \glspl{ssm} discussed above, we found a line in the \gls{uim} vertical to \gls{tst} longitudinal transfer function at 431.34 Hz, which was not present in the 2012 suspension model (Figure \ref{fig:v2l}). As above, we can examine the motion of the ANSYS model at this frequency when the \gls{uim} nodes are driven in the vertical direction. Again we see distortions in the areas of the \gls{uim} base plate under the spring supports, though in this case the motion between front and back is in-phase (Figure \ref{fig:uim_def_431}). The measured thickness of the \gls{uim} base plate shows a similar pole/zero combination to the \gls{ssm}, though again at a lower frequency (Figure \ref{fig:v2lmeas}).

As with the first line investigated, this one appears to be the result of distortions to the \gls{uim} directly under the supports holding the \gls{pum} and \gls{tst}. Without further investigation of the pendulum's response behavior, we cannot say for sure that the line found  in \gls{fea} is equivalent, but this model can give a starting point for attempts to reduce noise sources.

\section{Conclusions}
Any future \gls{gw} detector is likely to exhibit unintended mechanical modes in the suspension system that generate spectral lines in the strain sensitivity. Planned next-generation detectors such as Cosmic Explorer \cite{ce} and Einstein Telescope \cite{et} will use similar suspension designs with larger masses, potentially increasing the stresses induced on the upper stages, and so effects like the ones examined here should be taken into account. By modeling the design in \gls{fea}, we may be able to foresee these lines, and either shift them to a less valuable frequency or reduce them significantly. While \glspl{ssm} are useful for simulating the behavior of a particular suspension, using \gls{fea} can inform the reasons for that behavior. Models like the one shown here can illustrate internal modes, and identify cross-couplings. With this more detailed but still virtual view of the suspensions, we can assess the model for unexpected noise sources, and resolve them before finalizing and fabricating the design.

\section*{Acknowledgments}
The authors thank Giles Hammond, Alan Cumming, and Graeme Eddolls at the University of Glasgow for the use of their pendulum model.

This material is based upon work supported by the National Science Foundation under Grant No. PHY-2012021.

\end{document}